\documentclass[a4paper,12pt]{article}

\usepackage[english]{babel}
\usepackage[verbose=true,twoside,hmarginratio=3:4]{geometry}
\usepackage{fabdoc}

\newtheorem{lemma}{Lemma}[section]
\newtheorem{theorem}{Theorem}[section]

\newcommand{\be}{\begin{equation}}
\newcommand{\ee}{\end{equation}}
\newcommand{\bea}{\begin{eqnarray}}
\newcommand{\eea}{\end{eqnarray}}
\newcommand{\sh}{\sinh}

\newcommand{\prf}{\noindent{\bf Proof}\ }
\newcommand{\calL}{{\cal L}}

\begin{document}

\begin{center}

{\Large\bfseries Renormalization of Non-Commutative $\Phi^4_4$ Field Theory in $x$ Space}

\vskip 4ex

Razvan \textsc{Gurau}, Jacques \textsc{Magnen$^*$}, \\
Vincent \textsc{Rivasseau} and
Fabien \textsc{Vignes-Tourneret}

\vskip 3ex  

\textit{Laboratoire de Physique Th\'eorique, B\^at.\ 210, CNRS UMR 8627\\
Universit\'e Paris XI,  F-91405 Orsay Cedex, France\\
$^*$Centre de Physique Th\'eorique, CNRS UMR 7644\\
Ecole Polytechnique F-91128 Palaiseau Cedex, France}
\\
e-mail: \texttt{razvan.gurau@th.u-psud.fr}, 
\texttt{magnen@cpht.polytechnique.fr}, 
\texttt{vincent.rivasseau@th.u-psud.fr}, 
\texttt{fabien.vignes@th.u-psud.fr}
\end{center}

\vskip 5ex

\begin{abstract}
In this paper we provide a new proof that the Grosse-Wulkenhaar
non-commutative scalar $\Phi^4_4$ theory
is renormalizable to all orders in perturbation theory, and extend it to more
general models with covariant derivatives. 
Our proof relies solely on 
a multiscale analysis in $x$ space.
We think this proof is simpler and could be more adapted to
the future study of these theories (in particular at the non-perturbative or constructive level). 
\end{abstract}

\section{Introduction}
\setcounter{equation}{0}

In this paper we recover the proof of perturbative 
renormalizability of non-com\-mutative $\Phi^4_4$ field theory
\cite{GrWu03-1,GrWu04-3,Rivasseau2005bh} by a method solely based on $x$ space.
In this way we avoid completely the sometimes tedious use of the matrix basis and 
of the associated special functions of \cite{GrWu03-1,GrWu04-3,Rivasseau2005bh}.
We also extend the corresponding BPHZ theorem to the 
more general complex Langmann-Szabo-Zarembo $\bar\varphi \star \varphi \star \bar\varphi \star \varphi $ model 
with covariant derivatives, hereafter called the LSZ model.
This model has a slightly more complicated propagator, and is exactly solvable in a certain limit \cite{Langmann:2003if}.

Our method builds upon previous work of Filk and Chepelev-Roiban
\cite{Filk:1996dm,Chepelev:2000hm}. These works however remained inconclusive \cite{CheRoi}, since these 
authors used the right interaction but not the right propagator, hence
the problem of ultraviolet/infrared mixing prevented them from obtaining a finite 
renormalized perturbation series. The Grosse Wulkenhaar breakthrough was to realize
that the right propagator in non-commutative field theory 
is not the ordinary commutative propagator, but has to be modified to obey Langmann-Szabo duality \cite{LaSz,GrWu04-3}.

Non-commutative field theories (for a general review see \cite{DouNe}) deserve a thorough 
and systematic investigation. Indeed they may be relevant for physics beyond the standard model.
They are certainly effective models for certain limits of string theory
\cite{a.connes98:noncom,Seiberg:1999vs}. Also they form almost certainly the
correct framework for a microscopic {\it ab initio} understanding 
of the quantum Hall effect which is currently lacking.
We think that $x$ space-methods are probably more powerful for the future 
systematic study of  the noncommutative Langmann-Szabo covariant field theories.

Fermionic theories such as the two dimensional Gross-Neveu model can be shown to be
renormalizable to all orders in their Langmann-Szabo covariant versions, using either the matrix basis
or the direct space version developed here \cite{RenNCGN05}. However the $x$-space version seems the most 
promising for a complete non perturbative construction, using Pauli's principle to controll the
apparent (fake) divergences of perturbation theory. 
In the case of $\phi^4_4$, recall that although the commutative version is until now fatally flawed 
due to the famous Landau ghost, there is some hope that the non-commutative field theory treated 
at the perturbative level in this paper may also exist at the constructive level \cite{GrWu04-2,Rivasseau:2004az}. 
Again the $x$-space formalism is probably better than
the matrix basis for a rigorous investigation of this question.

In the first section of this paper we establish the $x$-space power counting 
of the theory using the Mehler kernel form of the propagator in direct space given in \cite{toolbox05}.
In the second section we prove that the divergent subgraphs can be renormalized
by counterterms of the form of the initial Lagrangian.
The LSZ models are treated in the Appendix.

\paragraph{Acknowledgment}
We thank V.~Gayral and R.~Wulkenhaar for useful discussions on this work.

\section{Power Counting in $x$-Space}
\setcounter{equation}{0}

\subsection{Model, Notations}

The simplest noncommutative $\varphi^4_4$ theory is defined on ${\mathbb R}^4$ equipped
with the associative and noncommutative Moyal product
\begin{align}
  (a\star b)(x) &= \int \frac{d^4k}{(2\pi)^4} \int d^4 y \; a(x{+}\tfrac{1}{2}
  \theta {\cdot} k)\, b(x{+}y)\, \mathrm{e}^{\mathrm{i} k \cdot y}\;.
\label{starprod}
\end{align}

The renormalizable action functional introduced in \cite{GrWu04-3} is
\begin{equation}\label{action}
S[\varphi] = \int d^4x \Big( \frac{1}{2} \partial_\mu \varphi
\star \partial^\mu \varphi + \frac{\Omega^2}{2} (\tilde{x}_\mu \varphi )
\star (\tilde{x}^\mu \varphi ) + \frac{1}{2} \mu_0^2
\,\varphi \star \varphi 
+ \frac{\lambda}{4!} \varphi \star \varphi \star \varphi \star
\varphi\Big)(x)\;,
\end{equation}
where $\tilde{x}_\mu=2(\theta^{-1})_{\mu\nu} x^\nu$ and the Euclidean
metric is used.

In four dimensional $x$-space the propagator is \cite{toolbox05}
\begin{equation}
C(x,x')=\frac{\Omega ^2}{[2\pi\sh\Omega t ]^2}
e^{-\frac{\Omega\coth\Omega t}{2}(x^2+x'^2)-
\frac{\Omega}{\sh\Omega t}x \cdot x'    - \mu_0^2 t }
\end{equation}
and the (cyclically invariant) vertex is \cite{Filk:1996dm}
\begin{equation}\label{vertex}
V(x_1, x_2, x_3, x_4) = \delta(x_1 -x_2+x_3-x_4 )e^{i
\sum_{1 \les i<j \les 4}(-1)^{i+j+1}x_i \theta^{-1}  x_j}
\end{equation}
where we note\footnote{Of course two different $\theta$ parameters could be used for the two 
symplectic pairs of variables of  ${\mathbb R}^4$.}
$x \theta^{-1}  y  \equiv  \frac{2}{\theta} (x_1  y_2 -  x_2  y_1 +  x_3  y_4 - x_4  y_3 )$.

The main result of this paper is a new proof in configuration space of
\begin{theorem}[BPHZ Theorem for Noncommutative $\Phi^4_4$ \cite{GrWu04-3,Rivasseau2005bh}]
\label{BPHZ1}
The theory defined by the action (\ref{action}) 
is renormalizable to all orders of perturbation theory.
\end{theorem} 

Let $G$ be an arbitrary connected graph. The amplitude associated with this graph is 
(with selfexplaining notations):
\begin{eqnarray}
A_G&=&\int \prod_{v,i=1,...4} dx_{v,i} \prod_l dt_l   \nonumber  \\
&& \prod_v \left[ \delta(x_{v,1}-x_{v,2}+x_{v,3}-x_{v,4})e^{\imath
\sum_{i<j}(-1)^{i+j+1}x_{v,i}\theta^{-1} x_{v,j}} \right]
\nonumber  \\
&&  \prod_l 
\frac{\Omega^2}{[2\pi\sinh(\Omega t_l)]^2}e^{-\frac{\Omega}{2}\coth(\Omega 
t_l)(x_{v,i(l)}^{2}+x_{v',i'(l)}^{2})
+\frac{\Omega}{\sinh(\Omega t_l)}x_{v,i(l)} . x_{v',i'(l)}   - \mu_0^2 t_l}\; .
\label{amplitude}
\end{eqnarray} 

For each line $l$ of the graph joining positions $x_{v,i(l)}$ and $x_{v',i'(l)}$, 
we choose an orientation and we define 
the ``short" variable $u_l=x_{v,i(l)}-x_{v',i'(l)}$ and the ``long" variable $v_l=x_{v,i(l)}+x_{v',i'(l)}$. 
With these notations, defining $\Omega t_l=\alpha_l$, the propagators in our graph can be 
written as:
\begin{equation}\label{tanhyp}
\int\prod_l \frac{\Omega d\alpha_l}{[2\pi\sinh(\alpha_l)]^2}
e^{-\frac{\Omega}{4}\coth(\frac{\alpha_l}{2})u_l^2-
\frac{\Omega}{4}\tanh(\frac{\alpha_l}{2})v_l^2  - \frac{\mu_0^2}{\Omega} \alpha_l}\; .
\end{equation} 

\subsection{Orientation and Position Routing}

A rule to solve the $\delta$ functions at every vertex is a ``position routing"
exactly analog to a momentum routing in the ordinary commutative case, except for the
additional difficulty of the cyclic signs which impose to orient the lines.
It is well known that there is no canonical such routing but 
there is a routing associated to any choice of a spanning tree in $G$. 
Such a tree choice is also useful to orient the lines of the graph, 
hence to fix the exact sign definition of the ``short" line variables $u_l$, and to 
optimize the multiscale power counting bounds below.

\begin{figure}[htbp]
\centering
\includegraphics[scale=1]{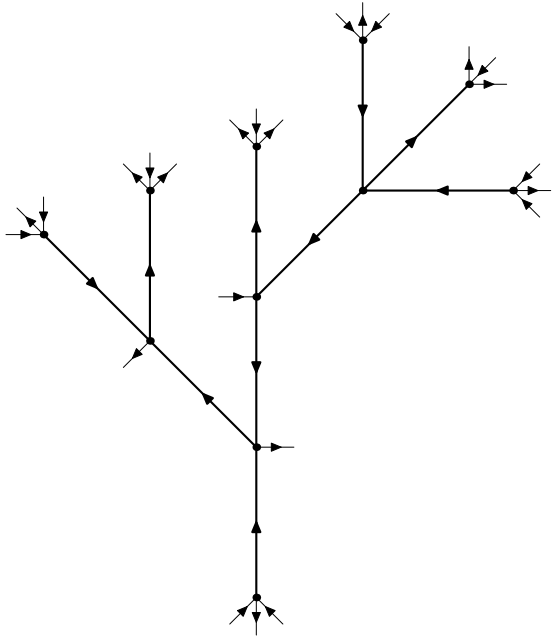}
\caption{Orientation of a tree}
\label{fig:treeorient}
\end{figure}

Let $n$ be the number of vertices of $G$, $N$ 
the number of its external fields, and $L$ the number of internal lines of $G$. We have $L= 2n-N/2$. 
Let $T$ be a rooted tree in the graph (when the graph is not a vacuum graph
it is convenient to choose for the root a vertex with external fields but this is not essential). 
We orient first all the lines of the tree and all the remaining 
half-loop lines or ``loop fields", following the cyclicity of the vertices.
This means that starting from an arbitrary orientation of a first field at the root and 
inductively climbing into the tree, at each vertex we follow the cyclic order to alternate entering 
and exiting lines as in Figure \ref{fig:treeorient}. 

Every line of the tree by definition of this orientation has one end exiting a vertex and an other
entering another one. This may not be true for the loop lines, which join two ``loop fields". 
Among these, some exit one vertex and enter another; they are called  
well-oriented. But others may enter or exit at both ends. These loop lines
are subsequently referred to as ``clashing lines". If there are no clashing lines, 
the graph is called orientable. If not, it is called non-orientable. 

We will see below that non-orientable graphs are irrelevant in the renormalization group sense. 
In fact they do not occur at all in some particular models such as
the LSZ model treated in the Appendix, or in the most natural noncommutative Gross-Neveu models
\cite{RenNCGN05}. 

For all the well-oriented lines (hence all tree propagators plus some of the 
loop propagators) we define in the natural way $u_l=x_{v,i(l)}-x_{v',i'(l)}$ if the 
line enters at $x_{v,i(l)}$ and exits from $x_{v',i'(l)}$. 
Finally we fix an additional (completely arbitrary) auxiliary orientation for all
the clashing loop lines, and fix in the same way $u_l=x_v-x_{v'}$ with respect to this auxiliary orientation.
 
It is also convenient to define the set of ``branches" associated to the rooted tree $T$.
There are $n-1$ such branches $b(l)$, one for each of the $n-1$ lines $l$ of the tree,
plus the full tree itself, called the root branch, and noted $b_0$.
Each such branch is made of the subgraph $G_b$ containing all the vertices ``above $l$"
in $T$, plus the tree lines and loop lines joining these vertices. It has also ``external fields" 
which are the true external fields hooked to $G_b$, plus the loop fields in $G_b$ for the
loops with one end (or ``field") inside and one end 
outside $G_b$, plus the upper end of the tree line $l$ itself to which $b$ is associated.
In the particular case of the root branch, $G_{b_0} = G$ and
the external fields for that branch are simply  all true external fields. 
We call $X_b$ the set of all external fields $f$ of $b$.

We can now describe the position routing associated to $T$.
There are $n$ $\delta$ functions in (\ref{amplitude}), hence $n$ linear equations for the $4n$ positions,
one for each vertex. The momentum routing associated to the tree $T$ solves this system by passing 
to another equivalent system of  $n$ linear equations, one for each branch of the tree.
This equivalent system is obtained by summing the arguments of the 
$\delta$ functions of the vertices in each branch. Obviously the Jacobian of this transformation is 1,
so we simply get another equivalent set of $n$ $\delta$ functions, one for each branch.

\begin{figure}[htbp]
  \centering
  \includegraphics[scale=0.8]{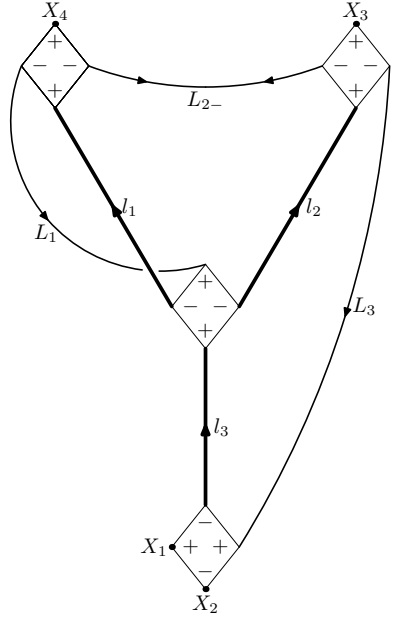}
  \caption{A branch}
  \label{fig:exbranch}
\end{figure}

Let us describe more precisely the positions summed in these branch equations, using the orientation.
Fix a particular branch $G_b$, with its subtree $T_b$. In the branch sum we find a sum 
over all the $u_l$ short parameters of the lines $l$ in $T_b$ and no $v_l$ long parameters
since $l$ both enters and exits the branch. This is also true for the set 
$L_b$ of well-oriented loops lines with both fields in the branch.
For the set $L_{b,+}$ of clashing loops lines with both fields entering the branch,
the short variable disappears and the long variable remains; the same is true but with a minus sign
for the set $L_{b,-}$ of clashing loops lines with both fields exiting the branch.
Finally we find the sum of positions of all external fields for the branch (with the signs
according to entrance or exit). For instance in the particular case of Figure \ref{fig:exbranch}, the delta function
is
\begin{equation}
\delta\lbt u_{l_{1}}+u_{l_{2}}+u_{l_{3}}+u_{L_{1}}+u_{L_{3}}-v_{L_{2}}+X_{1}-X_{2}+X_{3}+X_{4} 
\rbt \; .
\end{equation} 

The position routing is summarized  by:

\begin{lemma}[Position Routing]
We have, calling $I_G$ the remaining integrand in (\ref{amplitude}):
\begin{eqnarray}
A_G &=& \int \Big[ \prod_v  \big[ \delta(x_{v,1}-x_{v,2}+x_{v,3}-x_{v,4})\big] \, \Big]
I_G(\{x_{v,i}  \}  )   \\
&=& \int \prod_{b}
\delta \left(   \sum_{l\in T_b \cup L_b }u_{l} + \sum_{l\in L_{b,+}}v_{l}-\sum_{l\in L_{b,-}}v_{l}
+\sum_{f\in X_b}\epsilon(f) x_f \right) I_G(\{x_{v,i}  \}), \nonumber 
\end{eqnarray} 
where $\epsilon(f)$ is $\pm 1$ depending on whether the field $f$ enters or exits the branch.
\end{lemma}

Using the above equations one can at least solve all the long tree variables $v_l$ 
in terms of external variables, short variables and long loop variables, using the
$n-1$ non-root branches. There remains then the root branch $\delta$ function.
If $G_b$ is orientable, this $\delta$ function of branch $b_0$
contains only short and external variables, since $L_{b,+}$ and $ L_{b,-}$ are empty. 
If $G_b$ is non-orientable one can solve for an additional ``clashing" long loop variable.
We can summarize these observations in the following lemma:

\begin{lemma}\label{lemmarouting}
The position routing solves any long tree variable $v_l$ as a function 
of:
\begin{itemize}
\item the short tree variable $u_l$ of the line $l$ itself,
\item the short tree and loop variables with both ends in $G_{b(l)}$,
\item the long loop variables of the clashing loops with both ends in $G_{b(l)}$ (if any),
\item the short and long variables of the loop lines 
with one end inside $G_{b(l)}$ and the other outside.
\item the true external variables $x$ hooked to  $G_{b(l)}$.
\end{itemize}
The last equation corresponding to the root branch is particular. 
In the orientable case it does not contain any long variable, 
but gives a linear relation among the short variables and the external positions.
In the non-orientable case it gives a linear relation between 
the long variables $w$ of all the clashing loops in the graph 
some short variables $u$'s and all the external positions.
\end{lemma}

{}From now on, each time we use this lemma to solve the long tree variables $v_l$ 
in terms of the other variables, we shall call $w_l$ rather than $v_l$ the remaining $n+1 - N/2$ 
independent long loop variables. Hence looking at the long variables names the reader can check
whether Lemma \ref{lemmarouting} has been used or not.
\label{filkreduc1}

\subsection{Multiscale Analysis and Crude Power Counting}

In this section we follow the standard procedure of multiscale analysis \cite{Riv1}.
First the parametric integral for the propagator is sliced in the usual way :
\begin{equation}
C(u,v)=C^0(u,v)+\sum_{i=1}^{\infty}C^i(u,v),
\end{equation} 
with
\begin{equation}
C^i(u,v)=\int_{M^{-2i}}^{M^{-2(i-1)}}\frac{\Omega d\alpha} 
{[2\pi\sinh\alpha]^{2}}e^{-\frac{\Omega}{4}\coth\frac{\alpha}{2} 
u^2-\frac{\Omega}{4}\tanh\frac{\alpha}{2}v^2  -\frac{\mu_0^2}{\Omega} \alpha_l }
\end{equation}  

\begin{lemma} For some constants $K$ (large) and $c$ (small):
\begin{equation}\label{eq:propbound-phi4}
C^i (u,v) \les K M^{2i}e^{-c [ M^{i}\Vert u \Vert + M^{-i}\Vert v\Vert ] }
\end{equation} 
(which a posteriori justifies the terminology of ``long" and ```short" variables).
\end{lemma}
The proof is elementary, as it 
relies only on second order approximation of the hyperbolic functions near the origin.

Taking absolute values, hence neglecting all oscillations, leads to the following crude bound:
\begin{equation}\label{assignsum}
\vert A_G \vert \les \sum_{\mu}\int du_ldv_l\prod_l  C^{i_l}(u_l,v_l)
\prod_v \delta_v \ ,
\end{equation}
where $\mu$ is the standard assignment of an integer index $i_l$ to each propagator 
of each internal line $l$ of the graph $G$, which represents its ``scale". We will 
consider only amputated graphs. Therefore we have no 
external propagators, but only external vertices of the graph; 
in the renormalization group spirit, the convenient convention is to assign 
all external indices of these external fields to a  fictitious $-1$ ``background" scale. 

To any assignment $\mu$ and scale $i$
are associated the standard connected components $G_k^i$, $k=1,... ,k(i)$ 
of the subgraph $G^i$ made of all lines with scales $j\ges i$. These tree components 
are partially ordered according to their inclusion relations and the (abstract) tree
describing these inclusion relations is called the Gallavotti-Nicol\`o tree \cite{GaN}; 
its nodes are the $G_k^i$'s and its root is the complete graph $G$ (see Figure
\ref{multiscale-tools}).
\begin{figure}[htbp]
  \centering
  \subfloat[A $\varphi^{4}$ graph]{\label{fig:exgraph}\includegraphics[scale=0.9]{xphi4-fig.3}}
  \subfloat[Example of scale
  attribution]{\label{fig:exscale}\includegraphics[scale=0.7]{xphi4-fig.4}}\\
  \vspace*{1cm}
  \subfloat[The ``Gallavotti-Nicol\`o'' tree]{\label{fig:GNtree}\includegraphics[scale=1]{xphi4-fig.5}}
  \caption{}
  \label{multiscale-tools}
\end{figure}

More precisely for an arbitrary subgraph $g$ one defines: 
\begin{equation}
 i_g(\mu)=\inf_{l\in g}i_l(\mu) \quad , \quad 
 e_g(\mu)=\sup_{l \mathrm{~external~line~of~} g}i_l(\mu)\ .
\end{equation}
The subgraph $g$ is a $G_k^i$ for a given $\mu$ if and 
only if $i_g(\mu)\ges i> e_g(\mu)$. 
As is well known in the commutative field theory case, the key to optimize the bound over 
spatial integrations is to choose the real tree $T$ compatible with the abstract Gallavotti-Nicol\`o tree,
which means that the restriction $T_k^{i}$ of  $T$ to any $G_k^{i}$ must still span $G_k^{i}$.
This is always possible (by a simple induction from leaves to root).
We pick such a compatible tree $T$ and use it both to orient the graph as in the previous section and
to solve the associated branch system of $\delta$ functions according to Lemma \ref{lemmarouting}
We obtain: 
\begin{eqnarray}\label{bound1}
\vert A_{G,\mu} \vert &\les& K^n\prod_l M^{2i_l}\int du_ldv_l \prod_l
e^{-c [ M^{i_l}\Vert u_l \Vert  + M^{-i_l} \Vert v_l \Vert ]}
\prod_b\delta_b   \; .
\nonumber\\
&\les&  K^n\prod_l M^{2i_l}\int du_l dw_l \prod_l
e^{-c [ M^{i_l}\Vert u_l \Vert  + M^{-i_l} \Vert v_l (u,w,x) \Vert ]}  \delta_{b_0} \; .
\end{eqnarray} 

The key observation is to remark that any
long variable integrated at scale $i$ costs $KM^{4i}$ whereas any short 
variable integrated at scale $i$ brings $KM^{-4i}$, and the variables ``solved" by the $\delta$ functions
bring or cost nothing. For an orientable graph the optimal solution is easy:
we should solve the $n-1$ long variables $v_l$'s of the tree 
propagators in terms of the other variables, because this is the maximal number of long
variables that we can solve, and they have highest possible indices because
$T$ has been chosen compatible with the Gallavotti-Nicol\`o tree structure.
Finally  we still have the last $\delta_{b_0}$ function (equivalent to the overall momentum conservation
in the commutative case). It is optimal to use it
to solve one external variable (if any ) in terms of all the 
short variables and the external ones. Since external variables are typically smeared
against unit scale test functions, this leaves power counting invariant\footnote{In the case
of a vacuum graph, there are no external variables and we must 
therefore use the last $\delta_{b_0}$ function to solve 
the lowest possible short variable in terms of all others. In this way,
we loose the $M^{-4i}$ factor for this short integration. This is why the power counting
of a vacuum graph at scale $i$ is not given by the usual formula $M^{(4-N)i}= M^{4i}$ below at $N=0$,
but is in $M^{8i}$, hence worse by $M^{4i}$. 
This is of course still much better than the commutative case, because in that case and
in the analog conditions, that is without a fixed internal point, vacuum graphs
would be worse than the others by an ... 
infinite factor, due to translation invariance! In any case 
vacuum graphs are absorbed in the normalization of the theory, 
hence play no role in the renormalization.}.

The non-orientable case is slightly more subtle.
We remarked that in this case the system of branch equations allows 
to solve $n$ long variables as a functions of all the others. Should we always choose these $n$ 
long variables as the $n-1$ long tree variables plus one long loop variable? 
This is {\it not always} the  optimal choice.
Indeed when several disjoint $G^i_k$ subgraphs are non-orientable it is better
to solve more long clashing loop variables, essentially one per disjoint non-orientable $G^i_k$,
because they spare higher costs than if tree lines were chosen instead.
We now describe the optimal procedure, using words rather than equations 
to facilitate the reader's understanding.

Let ${\cal C}$ be the set of all the clashing loop lines. Each clashing
loop line has a certain scale $i$, therefore belongs to one and only one 
$G^i_k$ and consequently to all $G^j_{k'}\supset G^i_k$. We now define the 
set $S$ of $n$ long variables to be solved via the $\delta$ functions. First we 
put in $S$ all the $n-1$ long tree variables $v_l$. Then we scan 
all the connected components $G^i_k$ starting from the leaves towards the root,
and we add a clashing line to $S$ each time some new non-orientable component $G^i_k$ appears.
We also remove $p-1$ tree lines from $S$ each time $p\ges 2$ 
non-orientable components merge into a single one. In the end we obtain a new set
$S$ of exactly $n$ long variables.

More precisely suppose some $G^i_k$ at scale $i$ is a ``non-orientable leaf",
which means that is contains some clashing lines at scale $i$ but none at scales $j>i$. We 
then choose one (arbitrary) such clashing line and put it in the set $S$. Once a clashing
line is added to $S$ in this way it is never removed and no
other clashing line is chosen in any of the $G^j_k$ at 
lower scales $j<i$ to which the chosen line belongs. (The reader should be 
aware that this process allows nevertheless several clashing lines of $S$ to belong to a single $G^i_k$,
provided they were added to different connected components 
at upper scales.) When $p\ges 2$ non-orientable components merge at scale $i$
into a single non-orientable $G^i_k$, we can find $p-1$ lines 
in the part of the tree $T^i_k$ joining them together,
(e.g. taking them among the first lines on the unique paths in $T$ from these $p$ components
towards the root) and remove them from $S$.

We see that if we have added in all $q$ clashing lines to the set $S$, we 
have eliminated $q-1$ tree lines. The final set $S$ thus obtained in the end has 
exactly $n$ elements. The non trivial statement is that thanks to inductive use of Lemma \ref{lemmarouting}
in each $G^i_k$, we can solve all the long variables in the set $S$ with the branch system 
of $\delta$ functions associated to $T$. 

We perform now all remaining integrations. This spares the 
corresponding $M^{4i}$ integration cost for each long variable in $S$. 
For any line not in $S$ we see that the net power counting is 1, since the cost of the long variable
integration exactly compensates the gain of the short variable integration. 
But for any line in $S$ we earn the $M^{-4i}$ power counting of the 
corresponding short variable $u$ without paying the $M^{4i}$ cost of the long variable.

Gathering all the corresponding factors together 
with the propagators prefactors $M^{2i}$ leads to the 
following bound:
\begin{equation}
\vert A_{G,\mu}\vert  \  \les \  K^n
\prod_l M^{2i_l}\prod_{l\in S}M^{-4i_l } \ .
\end{equation} 
Remark that if the graph is well-oriented this formula remains true but the set $S$ consists of only 
the $n-1$ tree lines.

In the usual way of \cite{Riv1} we write 
\begin{equation}
\prod_{l}M^{2i_l}=\prod_{l}\prod_{i=1}^{i_l}M^2=
\prod_{i,k}\prod_{l\in G^i_k}M^2=\prod_{i,k}M^{2l(G^i_k)}
\end{equation} 
and
\begin{equation}
\prod_{l\in S}\prod_{i=1}^{i_l}M^{-4i_l}=
\prod_{i,k}\prod_{l\in G^i_k\cap S}M^{-4} ,
\end{equation} 
and we must now only count the number of elements in $G^i_k\cap S$.

If $G^i_k$ is orientable, it contains no clashing lines, hence 
$G^i_k\cap S=T^i_k$, and the cardinal of $T^i_k$ is $n(G^i_k)-1$. 

If $G^i_k$ contains one or more clashing lines and
$p$ clashing lines $l_1$, ... , $l_p$ in $G^i_k$ have been chosen to belong to $S$, then 
$p-1$ tree variables in $T^i_k$ have also been removed from $S$ and 
$G^i_k\cap S=T^i_k\cup\{l_1, ... \; , l_p\}-\{\mathrm{p-1~tree~variables}\}$, hence
the cardinal of $G^i_k\cap S$ is $n(G^i_k)$.

Using the fact that $2l(G^i_k)-4n(G^i_k)=-N(G^i_k)$ we can summarize 
these results in the following lemma:

\begin{lemma}\label{crudelemma}
The following bound holds for a connected graph (with external arguments integrated
against fixed smooth test functions):
\begin{equation}
\vert A_{G,\mu} \vert \les  K^n \prod_{i,k}M^{-\omega(G^i_k)}
\end{equation} 
for some (large) constant $K$, with $\omega(G^i_k)=N(G^i_k)-4$ if $G^i_k$ 
is orientable and $\omega(G^i_k)=N(G^i_k)$ if $G^i_k$ is non-orientable.
\end{lemma}
This lemma is optimal {\it if vertices oscillations are not taken into account}, and proves that 
non-orientable subgraphs are irrelevant.
But it is not yet sufficient for a renormalization theorem to all orders of perturbation. 

\subsection{Improved Power Counting}

Recall that for any non-commutative Feynman graph $G$ we can define
the genus of the graph, called $g$ and the number of faces ``broken by external legs", called $B$
\cite{GrWu04-3,Rivasseau2005bh}.
We have $g \ges 0$ and $B\ges 1$. 
The power counting established with the matrix basis in \cite{GrWu04-3,Rivasseau2005bh}, rewritten in the language
of this paper \footnote{Beware that the factor $i$ in \cite{Rivasseau2005bh} is now $2i$, and that the $\omega$
used here is the convergence rather than divergence degree. Hence there is both a sign change and a factor 2
of difference between the $\omega$'s of this paper and the ones of \cite{Rivasseau2005bh}.} is:
\begin{equation}\label{truepowercounting}
\omega (G) = N -4  + 8 g  + 4(B-1) \ ,
\end{equation}
hence we must (and can) renormalize only 2 and 4 point subgraphs
with $g=0$ and $B=1$, which we call \emph{planar regular}. They are the only non-vacuum graphs with $\omega \les 0$.

In the previous section we established that
\begin{equation}
\omega (G) \ges  N -4 \ , \ {\rm if}\   G\  {\rm orientable}\ , \ \   \omega (G) \ges  N  \ , \ 
{\rm if}\    G  \ {\rm non\ orientable}\ .
\end{equation} 

It is easy to check that planar regular subgraphs are orientable, but the 
converse is not true. Hence to prove that {\it orientable non-planar} 
subgraphs or {\it orientable planar} subgraphs with $B\ges 2$ are irrelevant 
requires to use a bit of the vertices oscillations to improve Lemma \ref{crudelemma}
and get:

\begin{lemma}\label{improvedbound}
For orientable subgraphs with $g\ges 1$ we have
\begin{equation}\label{improvednonplanar}
\omega (G) \ges  N + 4 \; .  
\end{equation} 
For orientable subgraphs with $g = 1$ and $B\ges 2$ we have
\begin{equation}\label{improvedbrokenfaces}
\omega (G) \ges  N  \; .
\end{equation} 
\end{lemma}
This lemma although still not giving (\ref{truepowercounting}) is sufficient for the purpose of this paper. 
For instance it implies directly that graphs which contain only irrelevant subgraphs in the sense of (\ref{truepowercounting}) have finite amplitudes uniformly bounded by $K^n$,
using the standard method of \cite{Riv1} to bound the assignment sum over $\mu$ in (\ref{assignsum}).

The rest of this subsection is essentially devoted to the proof of this Lemma \ref{improvedbound}.

We return before solving $\delta$ functions, hence to the $v$ variables. 
We will need only to compute in a precise way the oscillations which are 
quadratic in the long variables $v$'s to prove (\ref{improvednonplanar})
and the linear oscillations in $v \theta^{-1} x$ to prove (\ref{improvedbrokenfaces}). 
Fortunately an analog problem 
was solved in momentum space by Filk and Chepelev-Roiban \cite{Filk:1996dm,Chepelev:2000hm},
and we need only a slight adaptation of their work to position space.
In fact in this subsection short variables are quite inessential but it is convenient 
to treat on the same footing the long $v$  and the external $x$ variables, so we introduce a new 
global notation $y$ for all these variables. The vertices rewrite as
\begin{equation}
\prod_v\delta(y_1-y_2+y_3-y_4+\epsilon^iu_i)
e^{\imath \big(\sum_{i<j}(-1)^{i+j+1}y_i\theta^{-1}y_j + yQu + uRu \big)} \ .
\end{equation} 
for some inessential signs $\epsilon^i$ and some symplectic
matrices $Q$ and $R$.

Since we are not interested in the precise oscillations in the short $u$ variables 
we will note in the sequel quite sloppily $E_u$ any linear combination of 
the $u$ variables. Let's consider the first Filk reduction \cite{Filk:1996dm}, which contracts tree lines
of the graph. It creates progressively generalized vertices
with even number of fields. At a given induction step and for a tree line joining two 
such generalized vertices with respectively $p$ and $q-p+1$
fields ($p$ is even and $q$ is odd), 
we assume by induction that the two vertices are
\begin{eqnarray}\label{firstfilk1}
&& \delta(y_1-y_2+y_3...-y_p+E_u)
\delta(y_p-y_{p+1}+...-y_q+E_u)
\\
&& \hskip-1cm e^{  \imath \big(\sum_{1\les i<j\les p} (-1)^{i+j+1}y_i\theta^{-1}y_j+
\sum_{p\les i<j\les q} (-1)^{i+j+1}y_i\theta^{-1}y_j+yQu+ uRu   \big) } \ .\nonumber
\end{eqnarray} 
Using the second $\delta$ function we see that:
\begin{equation}\label{solveyp}
y_p=y_{p+1}-y_{p+2}+....+y_q-E_u \ .
\end{equation} 
Substituting this expression in the first $\delta$ function we get:
\begin{eqnarray}\label{firstfilk2}
  &&\delta(y_1-y_2+...-y_{p+1}+..-y_q+E_u)
  \delta(y_p-y_{p+1}+...-y_q+E_u)
 \\
 && \hskip-1cm e^{\imath\big( \sum_{1\les i<j\les p} (-1)^{i+j+1}y_i\theta^{-1}y_j+
\sum_{p\les i<j \les q} (-1)^{i+j+1}y_i\theta^{-1}y_j+yQu+ uRu \big)} \ . \nonumber
\end{eqnarray} 
 
The quadratic terms which include $y_p$ in the exponential are (taking 
into account that $p$ is an even number):
\begin{equation}
\sum_{i=1}^{p-1}(-1)^{i+1}y_i\theta^{-1}y_p+\sum_{j=p+1}^q
  (-1)^{j+1}y_p\theta^{-1}y_j \ .
\end{equation} 
Using the expression (\ref{solveyp}) for $y_p$ we see that the second term gives only 
terms in $yLu$. The first term yields:
\begin{equation}
\sum_{i=1}^{p-1}\sum_{j=p+1}^q (-1)^{i+1+j+1}y_i\theta^{-1}y_j=
  \sum_{i=1}^{p-1}\sum_{k=p}^{q-1}(-1)^{i+k+1}y_i\theta^{-1}y_j \ ,
\end{equation} 
which reconstitutes the crossed terms, and we have recovered the inductive form
of the larger generalized vertex.
 
\begin{figure}[htbp]
\centering
\includegraphics[scale=1.3]{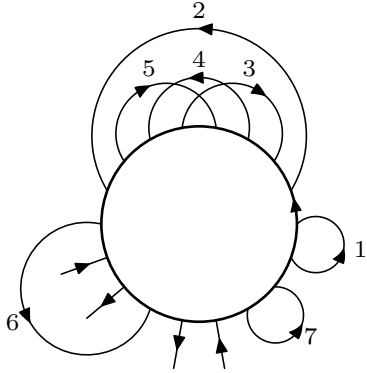}
\caption{A typical rosette}
\label{fig:exrosette}
\end{figure}

One should be aware that $y_p$ has disappeared from the final result, but 
that all the subsequent $y_{s>p}$ have changed sign.
This complication arises because of the cyclicity of the vertex.
As $p$ was chosen to be even (which implies $q$ odd)  
we see that $q-1$ is even as it should. Consequently by 
this procedure we will always treat only even vertices.
We finally rewrite the product of the two vertices as:
\begin{eqnarray}
&&\delta(y_1-y_2+...+y_{p-1}-y_{p+1}+..-y_q+E_u) \delta(y_p-y_{p-1}+...-y_q+E_u)
\nonumber\\
&&  e^{\imath\big( \sum_{1\les i<j\les q}(-1)^{i+j+1}y_i\theta^{-1}y_j+yQu+ uRu \big)}
\end{eqnarray} 
where the exponential is written in terms of the {\it reindexed} vertex 
variables. In this way we can contract all lines of a spanning tree $T$
and reduce $G$ to a single vertex with ``tadpole loops" called a ``rosette graph"
\cite{Chepelev:2000hm}.  
In this rosette to keep track of cyclicity is essential so rather than the ``point-like"
vertex of \cite{Chepelev:2000hm} we prefer to draw the rosette as a cycle 
(which is the border of the former tree) bearing loops lines on it 
(see Figure \ref{fig:exrosette}).
Remark that the rosette can also be considered as a big vertex, with $r=2n+2$ fields, on which $N$
are external fields with external variables $x$ and $2n+2-N$ are loop fields for the corresponding
$n+1-N/2$ loops. When the graph is orientable (which is the case to consider in Lemma
\ref{improvedbound}, 
the fields alternatively enter and exit, and correspond
to the fields on the border of the tree $T$, which we meet turning around counterclockwise
in Figure \ref{fig:treeorient}.
In the rosette the long variables $y_l$
for $l$ in $T$ have disappeared. Let us call $z$ the set of remaining long loop and external variables. Then the
rosette vertex factor is
\begin{equation}\label{rosettefactor}
\delta(z_1-z_2+...-z_r+E_u)
e^{\imath\big(\sum_{1\les i<j\les r}(-1)^{i+j+1}z_i\theta^{-1}z_j+zQu+ uRu\big)} \ .
\end{equation} 

The initial product of $\delta$ functions has not disappeared
so we can still write it as a product over branches like in the previous section and use it to solve
the $y_l$ variables in terms of the $z$ variables and the short $u$ variables. The net effect
of the Filk first reduction was simply to rewrite the
root branch $\delta$ function and the combination of all vertices oscillations 
(using the other $\delta$ functions) as the new big vertex or rosette factor
(\ref{rosettefactor}).

The second Filk reduction \cite{Filk:1996dm} further simplifies the rosette factor by erasing the
loops of the rosette which do not cross any other loops or arch over external fields. 
Here again the same operation is possible. 
Consider indeed such a rosette loop $l$
(for instance loop 2 in Figure \ref{fig:exrosette}). This means that
on the rosette cycle there is an even number of vertices in betwen the two ends of that loop and 
moreover that the sum of $z$'s in betwen these two ends must be zero, since they are loop variables
which both enter and exit between these ends. 
Putting together all the terms in the exponential which contain $z_l$ we 
conclude exactly as in \cite{Filk:1996dm} that these long $z$ variables 
completely disappears from the rosette oscillation factor, which simplifies as in \cite{Chepelev:2000hm}
to 
\begin{equation}\label{rosettefactorsimp}
\delta(z_1-z_2+...-z_r+E_u)
e^{\imath\lbt z {\cal I}z+zQu+ uRu\rbt} \;  ,
\end{equation} 
where ${\cal I}_{ij}$ is the antisymmetric ``intersection matrix" of \cite{Chepelev:2000hm} (up to a different sign convention). Here ${\cal I}_{ij}= +1$ if oriented loop line $i$ crosses oriented loop line $j$  coming from its right,
${\cal I}_{ij} = -1$ if $i$ crosses $j$  coming from its left,
and ${\cal I}_{ij} = 0$ if $i$ and $j$ do not cross. These formulas are also true for
$i$ external line and $j$ loop line or the converse, provided one extends the
external lines from the rosette circle radially to infinity to
see their crossing with the loops. Finally when $i$ and $j$ are external lines
one should define ${\cal I}_{ij} = (-1)^{p+q+1}$ if $p$ and  $q$ are the numbering
of the lines on the rosette cycle (starting from an arbitrary origin).

If a node $G^i_{k}$ of the Gallavotti-Nicol\`o tree is orientable but non-planar ($g \ges 1$), 
there must therefore exist two intersecting loop lines in the rosette corresponding 
to this $G^i_k$, with long variables $w_1$ and $w_2$. 
Moreover since $G^i_{k}$ is orientable, none of the long loop variables associated 
with these two lines belongs to the set $S$ of long variables eliminated by the $\delta$
constraints. Therefore, after 
integrating the variables in $S$ the basic mechanism to improve the
power counting of a single non planar subgraph is the following:
\begin{eqnarray}\label{gainoscill}
&&\int dw_1dw_2 e^{-M^{-2i_1}w_1^2-M^{-2i_2}w_2^2
- iw_1\theta^{-1}w_2+w_1 . E_1(x,u)+w_2 E_2(x,u)}
\nonumber\\
&=& \int dw'_1dw'_2 e^{-M^{-2i_1}(w_1')^2
-M^{-2i_2}(w'_2)^2 +iw'_1\theta^{-1}w'_2 + (u,x)Q(u,x)}
\nonumber\\
&=&  K  M^{4i_1} \int dw'_2
e^{- (M^{2i_1}+ M^{-2i_2})(w'_2)^2 }=
K M^{4i_1}M^{-4i_2} \; .
\end{eqnarray} 
In these equations we used for simplicity $M^{-2i}$ 
instead of the correct but more complicated factor $(\Omega /4) \tanh (\alpha /2 )$
(see \ref{tanhyp}) (of course this does not change the argument) and we performed
a unitary linear change of variables $w'_1 = w_1 + \ell_1 (x, u)$, $w'_2 = w_2 + \ell_2 (x, u)$
to compute the oscillating $w'_1$ integral. The gain in (\ref{gainoscill}) is
$M^{-8i_2}$, which is the difference between $M^{-4i_2}$ and 
the normal factor $M^{4i_2}$ that the $w_2$ integral would have cost if
we had done it with the regular $e^{-M^{-2i_2}w_2^2}$ factor for long variables. 
To maximize this gain we can assume $i_1 \les i_2$.

This basic argument must then be generalized to each non-planar leaf in the Gallavotti-Nicol\`o
tree. This is done exactly in the same way as the inductive definition of the set $A$
of clashing lines in the non-orientable case.
In any orientable non-planar `primitive" $G^i_k$ node (i.e. not containing sub non-planar nodes)
we can choose an arbitrary pair of crossing loop lines 
which will be integrated as in (\ref{gainoscill}) using this oscillation.
The corresponding improvements are independent.

This leads to an improved amplitude bound:
\begin{equation}
\vert A_{G,\mu} \vert \les K^n \prod_{i,k}M^{-\omega(G^i_k)}\ ,
\end{equation} 
where now $\omega(G^i_k)=N(G^i_k) + 4$ if $G^i_k$ 
is orientable and non planar (i.e. $g \ges 1$).
This bound proves (\ref{improvednonplanar}).

Finally it remains to consider the case of nodes $G^i_k$ which are planar orientable
but with $B \ges 2$. In that case there are no crossing loops in the rosette but
there must be at least one loop line arching over a non trivial subset
of external legs in the $G^i_k$ rosette (see line 6 in Figure \ref{fig:exrosette}). We have then a non trivial integration
over at least one external variable, called $x$, of at least one long loop variable called $w$.
This ``external" $x$ variable without the oscillation improvement 
would be integrated with a test function of scale 1 (if it is a true external line of scale $1$)
or better (if it is a higher long loop variable)\footnote{Since the loop line arches 
over a non trivial (i.e. neither full nor empty) subset
of external legs of the rosette, the variable $x$ cannot be the full combination 
of external variables in the ``root" $\delta$ function.}. But we get now
\begin{eqnarray}\label{gainoscillb}
&&\int dx dw e^{-M^{-2i}w^2
- iw\theta^{-1}x  +w.E_1(x',u)}
\nonumber\\
&=&  K  M^{4i} \int dx 
e^{-M^{+2i} x^2 }=
K' \ ,
\end{eqnarray} 
so that a factor $M^{4i}$ in the former bound becomes $O(1)$ hence is improved by $M^{-4i}$.
This proves (\ref{improvedbrokenfaces}) hence completes the proof of Lemma \ref{improvedbound}.
\qed
\bigskip\\
This method could be generalized to get the true power counting (\ref{truepowercounting}).
One simply needs a better description of the rosette oscillating factors when $g$ or $B$ increase.
It is in fact possible to ``disentangle" the rosette by some kind of ``third Filk move". Indeed the
rank of the long variables quadratic oscillations is exactly the genus, and the rank of the linear 
term coupling these long variables to the external ones is exactly $B-1$. So one can through a unitary
change of variables on the long variables inductively disentangle adjacent crossing pairs 
of loops in the rosette. This means that it is possible to diagonalize the 
rosette symplectic form through explicit moves of the loops along the rosette. 
Once oscillations are factorized in this way, the single improvements shown in this section
generalize to one improvement of $M^{-8i}$ per genus and one improvement of $M^{-4i}$ per broken face.
In this way the exact power counting (\ref{truepowercounting}) should be
recovered by pure $x$-space techniques which never require the use of the matrix basis. 
This study is more technical and not really necessary for the BPHZ theorem proved in
this paper.

\section{Renormalization}
\setcounter{equation}{0}

In this section we need to consider only divergent subgraphs, namely the planar
two and four point subgraphs with a single external face ($g=0$, $B=1$, $N=2$ or 4).
We shall prove that they can be renormalized by appropriate counterterms of the form
of the initial Lagrangian. We compute first
the oscillating factors $Q$ and $R$ of the short variables in (\ref{rosettefactorsimp}) for these graphs. 
This is not truly necessary for what follows, but is a good exercise.

\subsection{The Oscillating Rosette Factor}
In this subsection we define another more precise representation
for the rosette factor obtained after applying the first Filk moves
to a graph of order $n$.  We rewrite in terms of $u_l$ and $v_l$
the coordinates of the ends of the tree lines $l$, $l=1,\dots,n-1$ (those contracted in the first Filk moves),
but keep as variables called $s_1,\dots,s_{2n+2}$ the positions of all external fields and all ends of loop 
lines (those not contracted in the first Filk moves). 

\begin{figure}[htbp]
  \centering
  \includegraphics[scale=1.5]{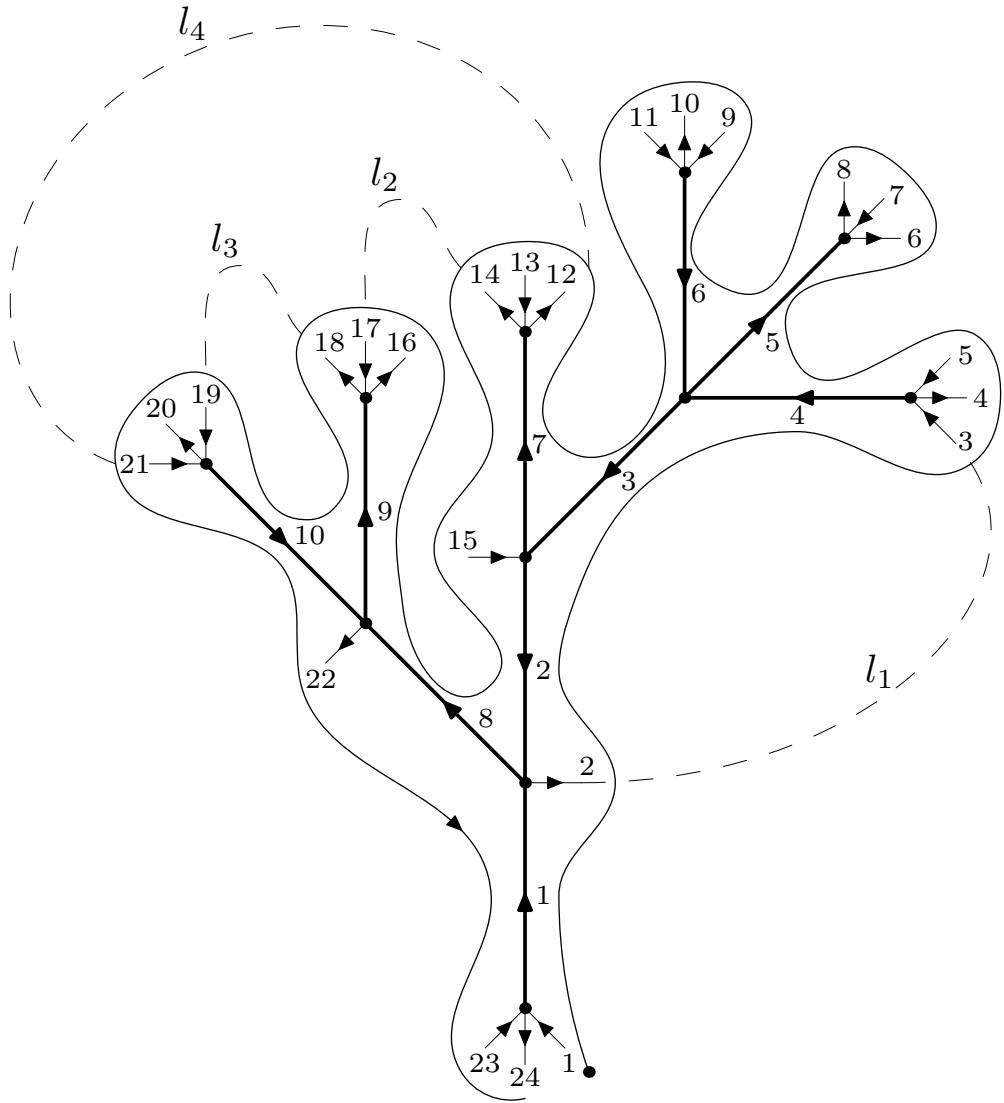}
  \caption{Total ordering of the tree lines and fields}
  \label{fig:turn-around-tree}
\end{figure}

We start from the root and turn around the tree in the trigonometrical sense. We number
separately all the fields as $1,\dots,2n+2$ and all the tree lines as $1,\dots,n-1$ in the order they are met,
but we also define a global ordering $\prec$ on the set of all the fields and
tree lines according to the order in which they are met (see Figure
\ref{fig:turn-around-tree}). In this way we know whether field number $p$ is
met before or after tree line number $q$. For example, in Figure
\ref{fig:turn-around-tree}, field number $8\prec$ tree line number $6$.

\begin{lemma}
The rosette contribution after a complete first Filk reduction is exactly:
\begin{eqnarray}
&&  \delta(s_1-s_2+\dots-s_{2n+2}+\sum_{l\in T}u_l)
e^{i\sum_{0\les i<j\les r}(-1)^{i+j+1}s_i\theta^{-1} s_j}
\nonumber \\
&& e^{-i\sum_{l \prec l'}u_l\theta^{-1} u_{l'}}e^{-i\sum_l \epsilon(l)\frac{u_l\theta^{-1} v_l}{2}} 
e^{i\sum_{l,i \prec l}(-1)^{i} s_i\theta^{-1}u_l+i\sum_{l,i \succ l}u_l \theta^{-1}(-1)^{i} s_i}\ ,
\end{eqnarray}
where $\epsilon(l)$ is $-1$ if the tree line $l$ is oriented towards the root and $+1$ if it is not. 
\end{lemma}

\noindent{\bf Proof:}
We proceed by induction. We contract the tree lines according to their ordering. 
In this way, at any step $k$ we contract a generalized vertex with $2k+2$
external fields corresponding to the contraction of the $k-1$ first lines with 
a usual four-vertex with $r=4$, and obtain a new generalized vertex with $2k+4$
fields.

We suppose inductively that the generalized vertex has the above form and prove that it keeps
this form after the contraction. We denote the external coordinates of this vertex as $s_1,\dots,s_{2k+2}$ 
and the coordinates of the four-vertex as $t_1,\dots,t_4$. We contract the propagator $(s_p,t_q)$ with 
associated variables $v=s_p+t_q$ and $u=(-1)^{p+1}s_p+(-1)^{q+1}t_q$. We also 
note that, since the tree is orientable, $p+q$ is odd.

Adding the arguments of the two $\delta$ functions gives the global 
$\delta$ function. We have the two equations:
\begin{equation}
s_1-s_2+\dots-s_{2k+2}+\sum u_s=0 \quad, \quad t_1-t_2+t_3-t_4=0 \; .
\end{equation}
Using the invariance of the $t$ vertex we can always eliminate 
the contribution of $t_q$ in the phase factor. We therefore have:
\begin{eqnarray}
\varphi&=& [s_1-s_2+\dots+(-1)^{p}s_{p-1}]\theta^{-1}
(-1)^{p}s_p\nonumber\\
&&+(-1)^{p}s_p\theta^{-1}[(-1)^{p+2}s_{p+1}+\dots-s_{2k+2}]
\nonumber\\
&=& [s_1-s_2+\dots+(-1)^ps_{p-1}]\theta^{-1} [-u+(-1)^{q+1}t_q] \nonumber\\
 &&+[-u+(-1)^{q+1}t_q]\theta^{-1}[(-1)^{p+2}s_{p+1}+\dots.-s_{2k+2}] .
\end{eqnarray}
As $(-1)^{q+1}t_q=\sum_{i=1,i\neq q}^4(-1)^it_i$ we see that the $s\theta^{-1} t_q$ terms
in the above expression reproduce exactly the crossed terms needed to complete the first exponential.
We rewrite the other terms as:
\begin{eqnarray}
&&[s_1-s_2+\dots+(-1)^{p}s_{p-1}]\theta^{-1}(-u)+(-u)\theta^{-1} 
[(-1)^{p+2}s_{p+1}+\dots-s_{2k+2}] \nonumber \\
&=&[s_1-s_2+\dots+(-1)^{p}s_{p-1}]\theta^{-1}(-u)\nonumber\\
&&+(-u)\theta^{-1}[-s_1+s_2\dots+(-1)^p s_p-\sum_s u_s]\nonumber \\
&=&2[s_1-s_2+\dots+(-1)^{p}s_{p-1}]\theta^{-1}(-u)+(-u)\theta^{-1}(-1)^p s_p 
+u\theta^{-1}\sum_s u_s \nonumber\\
&=&2\sum_{i \prec l}(-1)^{i}s_i\theta^{-1}u+(-1)^{p+1}\frac{u\theta^{-1}v}{2}+
\sum_s u\theta^{-1}u_s\; .
\end{eqnarray}
where we have used $(-1)^p s_p=(-1)^p (v-u)/2$. 

Note that further contractions will not involve $s_1\dots s_{p-1}$. After collecting all the 
contractions and using the global delta function we write:
\begin{equation}
2 \sum_{l,i \prec l}(-1)^i s_i\theta^{-1}u_l=\sum_{l,i \prec  l}(-1)^is_i\theta^{-1}u_l+
\sum_{l,i \succ  l} u_l \theta^{-1} (-1)^i s_i+\sum_{l,l'}u_l\theta^{-1}u_{l'} ,
\end{equation}
and the last term is zero by the antisymmetry of $\theta^{-1}$.
\qed
\bigskip\\
We note $\calL$ the set of loop lines, and analyze now further the rosette contribution for planar graphs. We 
call now $x_{i},\,i=1,\dots, N$ the $N$ external positions. We choose as first external field $1$ 
an arbitrary entering external line. We define an ordering among the set of all lines,
writing $l' \prec l$ if both ends 
of $l'$ are before the first end of $l$ when turning around the tree as in
Figure \ref{fig:turn-around-tree} where $l_{1}\prec l_{2}$. Analogously we define 
$l \prec j$ when $j$ is an external vertex ($l_{1}\prec x_{4}$ in Figure
\ref{fig:turn-around-tree}). We define $l'\subset  l$ if both ends of $l'$ lie
in between the ends of $l$ on the rosette ($l_{2}\subset l_{4}$ in Figure \ref{fig:turn-around-tree}). 
We count a loop line as positive if it turns in the trigonometric sense like the rosette
and negative if it turns clockwise. Each loop line $l \in \calL$ has now a sign 
$\epsilon(l)$ associated with this convention, and we now explicit its end
variables in terms of $u_l$ and $w_l$.

With these conventions we prove the following lemma:
\begin{lemma}\label{exactoscill}
The vertex contribution for a planar regular graph is exactly:
\begin{eqnarray}
&&\delta(\sum_{i}(-1)^{i+1}x_{i}+\sum_{l\in T\cup \calL} u_l)
e^{\imath\sum_{i,j}(-1)^{i+j+1}x_{i}\theta^{-1} x_{j}} 
\nonumber \\ 
&&e^{\imath\sum_{l\in T \cup \calL,\;   l \prec j}u_l\theta^{-1} (-1)^{j}x_j
+\imath\sum_{l\in T \cup \calL,\; l \succ j }(-1)^j x_{j}\theta^{-1} u_l}
\nonumber \\
&&e^{-\imath\sum_{l,l'\in T \cup \calL,\; l \prec l' }u_l\theta^{-1} u_{l'}
-\imath\sum_{l\in  T}\frac{u_l\theta^{-1} v_l}{2}\epsilon(l)
-\imath\sum_{l\in \calL}\frac{u_l\theta^{-1} w_l}{2}\epsilon(l)}
\nonumber \\
&&e^{-\imath\sum_{l\in\calL,\, l' \in \calL \cup T;\; l'\subset  l}u_{l'}\theta^{-1} w_l \epsilon(l)} \ .
\end{eqnarray}  
\end{lemma}

\prf We see that the global root $\delta$ function has the argument:
\begin{equation}
\sum_i (-1)^{i+1}x_i+\sum_{l\in \calL\cup T}u_l .
\end{equation}
Since the graph has one broken face we always have an even number of vertices on the external face between 
two external fields. We express all the internal loop variables as 
functions of $u$'s and $w$'s. Therefore the quadratic term in the external 
vertices can be written as:
\begin{equation}
\sum_{i<j}(-1)^{i+j+1}x_i\theta^{-1} x_j\ .
\end{equation}

The linear term in the external vertices is:
\begin{align}
&\sum_{i<j}(-1)^{i+1}s_i\theta^{-1} (-1)^jx_j+\sum_{i>j}(-1)^jx_j\theta^{-1} 
(-1)^{i+1}s_i\nonumber\\
&+\sum_{l\in T, l \succ j}(-1)^{j}x_{j}\theta^{-1} u_l +\sum_{l\in T, l  \prec j} u_l \theta^{-1}(-1)^{j}x_{j}\nonumber\\
=&\sum_{l'\in \calL, l' \succ j}u_{l'}\theta^{-1} (-1)^jx_j+\sum_{l'\in \calL, l' \succ j}(-1)^j x_j \theta^{-1} u_{l'}
\nonumber\\
&+ \sum_{l\in T, l \succ j}(-1)^{j}x_{j}\theta^{-1} u_l+\sum_{l\in T, l \prec j} u_l \theta^{-1}(-1)^{j}x_{j}\ .
\end{align}

Consider a loop line from $s_p$ to $s_q$ with $p<q$. Its contribution to the 
vertex amplitude decomposes in a "loop-loop" term and a "loop-tree" term. The first one is:
\begin{align}
&\sum_{i<p}(-1)^{i+1}s_i\theta^{-1} (-1)^p s_p+\sum_{\substack{p<i\\i\neq q}}(-1)^ps_p\theta^{-1}
(-1)^{i+1}s_i+s_{p}\theta^{-1}s_{q}\nonumber\\
&+\sum_{\substack{i<q\\i\neq p}}(-1)^{i+1}s_i\theta^{-1} (-1)^q s_q+\sum_{q<i}(-1)^ps_q\theta^{-1}
(-1)^{i+1}s_i\nonumber\\
=&\sum_{i<p}(-1)^{i+1}s_i\theta^{-1} [(-1)^{p}s_p+(-1)^qs_q] \nonumber\\
&+\sum_{q<i}[(-1)^ps_p+(-1)^{q}s_q]\theta^{-1} (-1)^{i+1}s_i\nonumber\\
&+\sum_{p<i<q}(-1)^{i+1}s^i\theta^{-1} [(-1)^{p+1}s_p+(-1)^qs_q]+
s_p\theta^{-1} s_q \ .
\end{align}

Taking into account that $(-1)^{i+1}s_i+(-1)^{j+1}s_j=u_{l'}$ if $s_i$ and 
$s_j$ are the two ends of the loop line $l'$, we can rewrite the above 
expression as:
\begin{eqnarray}
\sum_{l' \prec l}u_{l'}\theta^{-1} (-u_l)+\sum_{l' \succ l}(-u_l)\theta^{-1} u_{l'}+
\sum_{l'\subset  l}u_{l'}\theta^{-1} (-1)^{p+1}w_l 
\nonumber\\
+ (-1)^{p+1}\frac{u_l\theta^{-1} w_l}{2}+\sum_{l',l\subset  l'}u_l\theta^{-1} 
(-1)^{i+1}w_{l'}\ ,
\end{eqnarray}
where $l$ is fixed in all the above expressions.
Summing the contributions of all the lines (being careful not to count the same term twice) we get the 
final result:
\begin{equation}
-\sum_{l'  \prec l}u_{l'}\theta^{-1} u_l-\sum_{l,l'\subset  l}u_{l'}\theta^{-1} 
w_l~\epsilon(l)-\sum_l\frac{u_l\theta^{-1} w_l~\epsilon(l)}{2}\ .
\end{equation}

We still have to add the "loop-tree" contribution. It reads:
\begin{align}
&\sum_{l'\in T,l'\prec p}u_{l'}\theta^{-1}(-1)^{p}s_{p}+\sum_{l'\in T,l' \succ p}(-1)^{p} s_p \theta^{-1}
u_{l'}\nonumber\\
&+\sum_{l'\in T,l'  \prec q}u_{l'}\theta^{-1}(-1)^{q}s_{q}+\sum_{l'\in T,l' \succ q}(-1)^{q} s_q \theta^{-1}u_{l'}\nonumber\\
=&\sum_{l'\in T;l'  \prec p,q}u_{l'}\theta^{-1}[(-1)^{p}s_p+(-1)^q s_q]+\sum_{l'\in T;l' \succ p,q} [(-1)^{p}s_p+(-1)^q s_q]\theta^{-1}u_{l'}\nonumber\\
&+\sum_{l'\in T;p\prec  l' \prec q}u_{l'}\theta^{-1}[(-1)^{p+1}s_p+(-1)^{q}s_q]\nonumber\\
=&\sum_{l'\in T;l'\prec l}u_{l'}\theta^{-1}(-u_l)+\sum_{l' \in T;l' \succ  l}(-u_l)\theta^{-1}u_{l'}+
\sum_{l'\in T;l'\subset  l}u_{l'}\theta^{-1}(-1)^{p+1}w_l\ .
\end{align}
Collecting all the factors proves the lemma.
\qed

\subsection{Renormalization of the Four-point Function}\label{ren4pt}

Consider a 4 point subgraph which needs to be renormalized,
hence is a node of the Gallavotti-Nicol\`o tree.
This means that there is $(i,k)$ such that $N(G^{i}_{k})=4$. The four external positions
of the amputated graph are labeled $x_{1},x_{2},x_{3}$ and $x_{4}$. We also define $Q$, $R$ and
$S$ as three skew-symmetric matrices of respective sizes $4\times l(G^{i}_{k})$,
$l(G^{i}_{k})\times l(G^{i}_{k})$
and $[n(G^{i}_{k})-1]\times l(G^{i}_{k})$, where we recall 
that $n(G)-1$ is the number of loops of a 4 point graph with $n$ vertices. The amplitude associated to the
connected component $G^{i}_{k}$ is then
\begin{eqnarray}
A(G^{i}_{k})(x_{1},x_{2},x_{3},x_{4})&=&\int
\prod_{\ell\in T^{i}_{k}}  du_{\ell}   C_{\ell}(x, u, w) 
\prod_{l \in G^{i}_{k},\,l\not\in T}  du_{l}  d w_{l} C_{l}(u_l, w_l) 
\nonumber\\
&&\hskip-4cm\delta\Big(x_{1}-x_{2}+x_{3}-x_{4}+\sum_{l\in G^{i}_{k}} u_l\Big)
e^{\imath \lbt \sum_{p<q}(-1)^{p+q+1}x_{p}\theta^{-1}
x_{q}+XQU+URU+USW \rbt } . \label{eq:4pt-ini}
\end{eqnarray}
The exact form of the factor $\sum_{p<q} (-1)^{p+q+1}x_{p}\theta^{-1}
x_{q}$  follows from Lemma \ref{exactoscill}. From this Lemma and (\ref{exactvvalue}) below 
would also follow exact expressions for
$Q$, $R$ and $S$, but we wont need them. The important fact is that 
there are no quadratic oscillations in $X$ times $W$ (because $B=1$) nor in $W$ times $W$ (because $g=0$).
$C_{l}$ is the propagator of the
line $l$. For loop lines $C_{l}$ is expressed in terms of $u_l$ and $w_l$
by formula (\ref{tanhyp}), (with $v$ replaced by our notation $w$ for long variables of loop lines). 
But for tree lines $\ell \in T^i_k$
recall that the solution of the system of branch $\delta$ functions for $T$ has reexpressed the
corresponding long variables $v_\ell$ in terms of the short variables $u$,
and the external and long loop variables of the branch graph $G_\ell$ 
which lies ``over" $\ell$ in the rooted tree $T$.
This is the essential content of the subsection \ref{filkreduc1}. 
More precisely consider a line $\ell \in T^i_k$ with scale $i(\ell)\ges i$;
we can write
\begin{equation}\label{exactvvalue}
v_\ell =  X_\ell + W_\ell + U_\ell 
\end{equation} 
where 
\begin{equation}\label{xvalue}
X_\ell = \sum_{e\in E(\ell)}  \epsilon_{\ell,e} x_{e}
\end{equation}
is a linear combination on the set of external variables of the branch graph $G_\ell$ with 
the correct alternating signs $\epsilon_{\ell,e}$, 
\begin{equation}\label{wvalue}
W_\ell = \sum_{l \in \calL (\ell)} 
\epsilon_{\ell,l} w_{l}
\end{equation}
is a linear combination over the set $\calL (\ell)$ of long loop variables 
for the external lines of $G_\ell$ (and $\epsilon_{\ell,l}$ are other signs), and
\begin{equation}\label{uvalue}
U_\ell = \sum_{l' \in S (\ell)} \epsilon_{\ell, l'} u_{l'}
\end{equation}
is a linear combination over a set $S_\ell$ 
of short variables that we do not need to know explicitly.
The tree propagator for line $\ell$ then is
\begin{equation}\label{propatre}
C_{\ell}(u_\ell, X_\ell, U_\ell, W_\ell) = \int_{M^{-2i(\ell)}}^{M^{-2(i(\ell)-1)}} 
\frac{\Omega d\alpha_\ell e^{-\frac{\Omega}{4} 
\{ \coth(\frac{\alpha_\ell}{2})u_l^2  + \tanh(\frac{\alpha_\ell}{2}) [X_\ell + W_\ell + U_\ell ]^2 \} }}{[2\pi\sinh(\alpha_\ell)]^2}
\ .
\end{equation}

To renormalize, let us call $e= \max e_p$, $p=1,...,4 $ the highest external index 
of the subgraph $G^{i}_{k}$.  We have $e<i$ since $G^{i}_{k}$
is a node of the Gallavotti-Nicol\`o tree. We evaluate $A(G^{i}_{k})$ on external 
fields\footnote{For the external index to be exactly $e$ the external smearing factor should be in fact 
$\prod_{p} \varphi^{\les e}(x_p) - \prod_{p} \varphi^{\les e-1}(x_p)$ but this subtlety is inessential.}
$\varphi^{\les e}(x_p)$ as:
\begin{eqnarray}
A(G^{i}_{k})&=&\int\prod_{p=1}^{4}dx_{p}\varphi^{\les e}(x_{p})\,
A(G^{i}_{k})(x_{1},x_{2},x_{3},x_{4})\nonumber\\
&=&\int\prod_{p=1}^{4}dx_{p } \varphi^{\les e}(x_{p})\ e^{\imath\text{Ext}}
\prod_{\ell\in T^{i}_{k} }  du_{\ell}   C_{\ell}(u_\ell, tX_\ell, U_\ell, W_\ell) 
\\
&&
\prod_{l \in G^{i}_{k} \, \ l \not \in T}  du_{l}  d w_{l} C_{l}(u_l, w_l) \ 
\delta\Big(\Delta+t\sum_{l\in G^{i}_{k}}u_{l}\Big)
e^{\imath tXQU+\imath URU+\imath USW}\  \Bigg|_{t=1}\ . \nonumber
\end{eqnarray}
with $\Delta = x_{1}-x_{2}+x_{3}-x_{4}$ and $\text{Ext}=\sum_{p<q=1}^{4}(-1)^{p+q+1}x_{p}\theta^{-1} x_{q}$. 

This formula is designed so that at $t=0$ all dependence on the external variables $x$
factorizes out of the $u,w$ integral in the desired vertex form for renormalization
of the $\varphi \star \varphi \star \varphi \star \varphi$ interaction in the action (\ref{action}).
We now perform a Taylor expansion to first order with respect to the $t$ variable and prove 
that the remainder term is irrelevant. Let $\mathfrak{U}=\sum_{l\in G^{i}_{k}}u_{l}$, and
\begin{eqnarray}
{\mathfrak R}(t) &=&   -\sum_{\ell\in T^{i}_{k} }\frac{\Omega}{4}\tanh(\frac{\alpha_\ell}{2}) \Bigg\{
t^2 X_\ell ^2 + 2t X_\ell \big[ W_\ell + U_\ell \big] \Bigg\} 
\nonumber\\
&\equiv& - t^2  {\cal A} X . X  -  2t {\cal A}X . (W + U )
\ .
\end{eqnarray} 
where ${\cal A}_\ell = \frac{\Omega}{4}\tanh(\frac{\alpha_\ell}{2})$, and $X\cdot Y$ means
$\sum_{\ell\in T^{i}_{k} } X_\ell  . Y_\ell $.
We have
\begin{eqnarray}
A(G^{i}_{k})&=&\int\prod_{p=1}^{4}dx_{p}\varphi^{\les e}(x_{p})\, e^{\imath\text{Ext}}
\prod_{\ell\in T^{i}_{k} }  du_{\ell}   \ C_{\ell}(u_\ell, U_\ell, W_\ell) 
\nonumber\\
&&
\bigg[ \prod_{l \in G^{i}_{k} \, \ l \not \in T}  du_{l}  d w_{l} C_{l}(u_l, w_l) \bigg]
\ e^{\imath URU+\imath USW}
\\
&&\hspace{-2cm}
\Bigg\{ \delta(\Delta)
+ \int_{0}^{1}dt\bigg[ \mathfrak{U}\cdot \nabla \delta(\Delta+t\mathfrak{U})
+\delta(\Delta+t\mathfrak{U}) [\imath XQU  + {\mathfrak R}' (t)]  \bigg] 
e^{\imath tXQU + {\mathfrak R}(t)}  \Bigg\} \ .\nonumber
\end{eqnarray}
where
$C_{\ell}(u_\ell, U_\ell, W_\ell) $ is given by (\ref{propatre})
but taken at $X_\ell=0$.

The first term, denoted by $\tau A$, is of the desired form (\ref{vertex}) times a number
independent of the external variables $x$. It is asymptotically constant in
the slice index $i$, hence the sum over $i$
at fixed $e$ is logarithmically divergent: this is the divergence
expected for the four-point function. It remains only to check that $(1-\tau)A$ converges
as $i-e \to \infty$.  But we have three types of terms in  $(1-\tau)A$, 
each providing a specific improvement over the regular, log-divergent power counting of $A$:

\begin{itemize}

\item The term $\mathfrak{U}\cdot \nabla \delta(\Delta+t\mathfrak{U})$. For this term, integrating by parts over
external variables, the $\nabla $ acts on external fields $\varphi^{\les e}$, 
hence brings at most $M^e$ to the bound,
whether the $\mathfrak{U}$ term brings at least $M^{-i}$.

\item  The term $XQU$. Here $X$ brings at most $M^e$ and $U$ brings at least $M^{-i}$.

\item The term ${\mathfrak R}' (t)$. It decomposes into terms in ${\cal A} X \cdot X$, ${\cal A} X\cdot U$
and ${\cal A} X\cdot W$. Here
the ${\cal A}_\ell$ brings at least $M^{-2 i(\ell)}$, $X$ brings at worst $M^{e}$,
$U$ brings at least $M^{-i}$ and $X_\ell  W_\ell$ brings at worst $M^{e+i(\ell)}$.
This last point is the only subtle one:
if $\ell \in T^i_k$, remark that because $T^i_k$ is a sub-tree within each
Gallavotti-Nicol\`o subnode of $G^i_k$, in particular all parameters
$w_{l'}$ for $l' \in\calL (\ell)$ which appear in $W_\ell$
must have indices lower or equal to $i(\ell)$ (otherwise they would have been chosen instead of $\ell$
in $T^i_k$).

\end{itemize}

In conclusion, since $i(\ell) \ges i$, the Taylor remainder term $(1-\tau)A$
improves the power-counting of the connected component
$G_{k}^{i}$ by a factor at least $M^{-(i-e)}$.
This additional $M^{-(i-e)}$ factor makes $(1-\tau)A(G^{i}_{k})$ convergent 
and irrelevant as desired. 

\subsection{Renormalization of the Two-point Function}\label{Ren2pt}

We consider now the nodes such that $N(G^{i}_{k})=2$.  We use the same notations than in the previous subsection.  The two external points are labeled $x$ and $y$. Using the global 
$\delta$ function,  which is now 
$\delta\Big(x-y + {\mathfrak U}\Big)$,
we remark that the external oscillation $e^{\imath x \theta^{-1} y}$
can be absorbed in a redefinition of the term $e^{\imath tXQU}$, which we do from now on.
Also we want to use expressions symmetrized over $x$ and $y$.
The full amplitude is
\begin{eqnarray}\label{2point1}
A(G^{i}_{k}) &=&\int dx dy 
\varphi^{\les e}(x)\varphi^{\les e}(y) \delta\Big(x-y + {\mathfrak U}\Big)
\\
&&  \prod_{l \in G^{i}_{k} ,\; l \not \in T}  du_{l}  d w_{l} C_{l}(u_l, w_l)
\nonumber\\
&&
\prod_{\ell\in T^{i}_{k} }  du_{\ell}   C_{\ell}(u_\ell, X_\ell, U_\ell, W_\ell) 
\ e^{\imath XQU+\imath URU+\imath USW}\; . \nonumber
\end{eqnarray}
First we write the identity
\begin{eqnarray}
\varphi^{\les e}(x)\varphi^{\les e}(y) &=& 
\frac 12 \bigg[ [\varphi^{\les e}(x)]^2 + [\varphi^{\les e}(y) ]^2 \; -\; 
[ \varphi^{\les e}(y) - \varphi^{\les e}(x)]^2 \bigg]\; ,
\label{eq:2pt-sym}
\end{eqnarray}  
we develop it as
\begin{eqnarray}\label{symdev}
\varphi^{\les e}(x)\varphi^{\les e}(y) &=& 
\frac 12 \Bigg\{ [\varphi^{\les e}(x)]^2 + [\varphi^{\les e}(y) ]^2 
- \bigg[   (y-x)^\mu \cdot \nabla_\mu \varphi^{\les e}(x)
\\&&
\hskip-1cm  + 
\int_{0}^{1}ds (1-s)  (y-x)^\mu (y-x)^\nu \nabla_\mu \nabla_\nu
\varphi^{\les e}(x + s(y-x))  \bigg]^2 \Bigg\}\; ,
\nonumber
\end{eqnarray}  
and substitute into (\ref{2point1}).
The first term $A_0$ is a symmetric combination 
with external fields at the same argument. Consider the case with the two external legs
at $x$, namely the term in $[\varphi^{\les e}(x)]^2 $. 
For this term we integrate over $y$. This uses the $\delta$ function. We 
perform then a Taylor expansion in $t$ at order $3$ of the  remaining function
\begin{equation}
  \label{eq:f}
  f(t)=  e^{\imath tXQ U   + {\mathfrak R}(t)}\; ,
\end{equation} 
where we recall that
${\mathfrak R}(t)= - [ t^2  {\cal A} X . X + 2t {\cal A}X . (W + U )]$. We get 
\begin{eqnarray}\label{2point2}
A_0 &=& \frac 12 \int dx  [\varphi^{\les e}(x)]^2    \,
e^{\imath  (URU+ USW)} 
\nonumber\\
&& \prod_{l \in G^{j}_{k} , \; l \not \in T}  du_{l}  d w_{l} C_{l}(u_l, w_l)
 \prod_{\ell\in T^{i}_{k}}  du_{\ell}   C_{\ell}(u_\ell, U_\ell, W_\ell) 
\nonumber\\
&& 
\lbt f(0)+f'(0)+\frac 12f''(0)+\frac 12\int_{0}^{1}dt\,(1-t)^{2}f^{(3)}(t)\rbt\
\ . 
\end{eqnarray}

In order to evaluate that expression, let $A_{0,0},A_{0,1},A_{0,2}$ be the
zeroth, first and second order terms in this Taylor
expansion, and $A_{0,R}$ be the remainder term. First,
\begin{eqnarray}
A_{0,0}&=&\int dx\, [\varphi^{\les e}(x)]^2  \; e^{\imath (URU+ USW)}
\prod_{l \in G^{i}_{k} \; , l \not \in T}  du_{l}  d w_{l} C_{l}(u_l, w_l)
\nonumber\\
&&
\prod_{\ell\in T^{i}_{k} }  du_{\ell}   C_{\ell}(u_\ell, U_\ell, W_\ell)  
\end{eqnarray}
is quadratically divergent and exactly of the expected form for the mass counterterm. Then
\begin{eqnarray}
A_{0,1}&=&
\frac 12 \int dx [\varphi^{\les e}(x)]^2 \ e^{\imath (URU+ USW)}
\prod_{l \in G^{i}_{k}  ,\; l \not \in T}  du_{l}  d w_{l} C_{l}(u_l, w_l)
\nonumber\\
&&
\prod_{\ell\in T^{i}_{k} }  du_{\ell}   C_{\ell}(u_\ell, U_\ell, W_\ell)     
\bigg(  \imath XQU  +  {\mathfrak R}' (0)   \bigg) 
\end{eqnarray} 
vanishes identically. Indeed all the terms are odd 
integrals over the $u,w$-variables. $A_{0,2}$ is more complicated:
\begin{eqnarray}\label{eqtwo2}
A_{0,2}&=& \frac 12 \int dx [\varphi^{\les e}(x)]^2 \ e^{\imath (URU+ USW)}
\prod_{l \in G^{i}_{k}  ,\; l \not \in T}  du_{l}  d w_{l} C_{l}(u_l, w_l)\nonumber\\
&&
\prod_{\ell\in T^{i}_{k} }  du_{\ell}   C_{\ell}(u_\ell, U_\ell, W_\ell)  \ \Bigg( -( XQU)^2   
\nonumber\\
&& \hskip -1cm- 4\imath XQU {\cal A}X \cdot (W + U )  -2 {\cal A}X \cdot X +4  [{\cal A}X \cdot (W + U )] ^2 
 \Bigg) .
\end{eqnarray}

The four terms in $(XQU)^2$, $XQU {\cal A}X \cdot W$ ${\cal A}X \cdot X$ and  $[{\cal A}X \cdot W ] ^2 $
are logarithmically divergent and contribute to the renormalization of the  harmonic frequency term
$\Omega$ in (\ref{action}). (The terms in $x^\mu x^\nu$ with  $\mu \ne \nu$ do not survive by parity
and the terms in $(x^\mu )^2$  have obviously the same coefficient. 
The other terms in $XQU {\cal A}X \cdot U$,
 $({\cal A}X \cdot U)({\cal A}X \cdot W)$ and  $[{\cal A}X \cdot U ] ^2 $ are irrelevant. Similarly the terms 
 in $A_{0,R}()x$ are all irrelevant.
 
For the term in $A_{0}(y)$ in which we have $\int dx [\varphi^{\les e}(y)]^2 $ we have to perform a similar
computation, but beware that it is now $x$ which is integrated with the $\delta$ function so that 
$Q$, $S$, $R$ and ${\mathfrak R}$ change, but not the conclusion.
 
Next we have to consider the term in $\bigg[ (y-x)^\mu \cdot \nabla_\mu \varphi^{\les e}(x) \bigg]^2 $
in (\ref{symdev}), for which we need to develop the
$f$ function only to first order. 
Integrating over $y$ replaces each $y-x$ by a ${\mathfrak U}$ factor so that we get a term
\begin{eqnarray}
A_{1}&=&  \frac 12\int dx\, \bigg[ {\mathfrak U}^\mu \cdot \nabla_\mu \varphi^{\les e}(x) \bigg]^2 
\; e^{\imath (URU+ USW)}\prod_{l \in G^{i}_{k} \; , l \not \in T}  du_{l}  d w_{l} C_{l}(u_l, w_l)
\nonumber\\
&&
\prod_{\ell\in T^{i}_{k} }  du_{\ell}   C_{\ell}(u_\ell, U_\ell, W_\ell) \bigg( f(0)+\int_0^1 dt  f'(t) dt \bigg)
\end{eqnarray}

The first term is
\begin{eqnarray}
A_{1,0} &=& \frac 12\int dx\, \bigg[ {\mathfrak U}^\mu \cdot \nabla_\mu \varphi^{\les e}(x) \bigg]^2 
\; e^{\imath (URU+ USW)}\prod_{l \in G^{i}_{k} \; , l \not \in T}  du_{l}  d w_{l} C_{l}(u_l, w_l)
\nonumber\\
&&
\prod_{\ell\in T^{i}_{k} }  du_{\ell}   C_{\ell}(u_\ell, U_\ell, W_\ell) 
\end{eqnarray}
The terms with $\mu \ne \nu$ do not survive by parity. The other ones reconstruct a counterterm proportional
to the Laplacian. The power-counting of this factor $A_{1,0}$ is improved, with respect to $A$, by
a factor $M^{-2(i-e)}$ which makes it only logarithmically divergent, as should 
be for a wave-function counterterm. 

The remainder term in $A_{1,R}^{x}$ has an additional factor at worst $M^{-(i-e)}$
coming from the $\int_0^1 dt  f'(t) dt $ term,
hence is irrelevant and convergent.

Finally the remainder terms $A_{R}$ with three or four gradients in (\ref{symdev})
are also irrelevant and convergent. Indeed  we have terms of various types:

\begin{itemize}

\item There are terms in $U^3$ with $\nabla^3 $. The  $\nabla $ 
act on the variables $x$, hence on external fields, hence bring at most $M^{3e}$ to the bound,
whether the $\mathfrak{U}^3$ brings at least $M^{-3i}$.

\item Finally there are terms with 4 gradients which are still smaller.
\end{itemize}

Therefore for the renormalized amplitude $A_{R}$
the power-counting is improved, with respect to $A_{0}$, by a factor $M^{-3(i-e)}$, 
and becomes convergent.

Putting together the results of the two previous section, we have
proved that the usual effective series  which expresses any connected function of the theory
in terms of an infinite set of effective couplings, related one to each other by a discretized flow \cite{Riv1},
have finite coefficients to all orders. Reexpressing these effective series in terms of the 
renormalized couplings would reintroduce in the usual way the Zimmermann's forests of ''useless" counterterms
and build the standard ``old-fashioned" renormalized series.
The most explicit way to check finiteness of these renormalized series in order to complete the ``BPHZ theorem"
is to use the standard ``classification of forests"  which distributes  Zimmermann's forests into packets
such that the sum over assignments in each packet is finite \cite{Riv1}\footnote{One could also use the popular
inductive scheme of Polchinski, which however does not extend yet to non-perturbative 
``constructive" renormalization}. This part is completely standard and identical
to the commutative case. Hence the proof of Theorem \ref{BPHZ1} is completed.

\appendix
\section{The LSZ Model}
\setcounter{equation}{0}

In this section we prove the perturbative renormalizability of a generalized\\
Langmann-Szabo-Zarembo model \cite{Langmann:2002ai}. It consists in a bosonic complex scalar field theory in
a fixed magnetic background plus an harmonic oscillator. The quartic interaction is of the Moyal type. The
action functional is given by 
\begin{align}
  S=&\int\frac 12\bar{\varphi}\lbt -D^{\mu}D_{\mu}+\Omega^{2}x^{2}+\mu_{0}^{2}\rbt\varphi
 +\lambda\,\bar{\varphi}\star\varphi\star\bar{\varphi}\star\varphi\label{eq:lszaction}
\end{align}
where $D_{\mu}=\partial_{\mu}-\imath B_{\mu\nu}x^{\nu}$ is the
covariant derivative. The $1/2$ factor is somewhat unusual in a complex
theory but it allows us to recover exactly the results given in
\cite{toolbox05} with $\Omega^{2}\rightarrow\omega^{2}=\Omega^{2}+B^{2}$.
By expanding the quadratic part of the action, we
get a $\Phi^{4}$-like kinetic part plus an angular momentum term:
\begin{align}
  \bar{\varphi}D^{\mu}D_{\mu}\varphi
  +\Omega^{2}x^{2}\bar{\varphi}\varphi=&\,\bar{\varphi}\lbt\Delta -\omega^{2}x^{2}-2BL_{5}\rbt\varphi
\end{align}
with $L_{5}=x^{1}p_{2}-x^{2}p_{1}+x^{3}p_{4}-x^{4}p_{3}= x \wedge \nabla$. Here the
skew-symmetric matrix $B$ has been put in its canonical form
\begin{equation}
  \label{eq:Bform}
  B=\begin{pmatrix}\begin{matrix}0&-1\\1&\phantom{-}0\end{matrix}&(0)\\
    (0)&\begin{matrix}0&-1\\1&\phantom{-}0\end{matrix}
    \end{pmatrix}.
\end{equation}
In $x$ space, the interaction term is exactly the same as (\ref{vertex}). The complex conjugation of the fields only selects the orientable graphs.\\
At $\Omega=0$, the model is similar to the Gross-Neveu theory. This will be
treated in a future paper \cite{RenNCGN05}. If we additionally set
$B=\theta^{-1}$ we recover the integrable LSZ model \cite{Langmann:2002ai}.

\subsection{Power Counting}
\label{sec:lszpowcount}

The propagator corresponding to the action (\ref{eq:lszaction}) has been
calculated in \cite{toolbox05} in the two-dimensional case. The
generalization to higher dimensions e.g. four, is straightforward:
\begin{align}
  C(x,y)=&\int_{0}^{\infty}dt\,\frac{\omega^{2}}{(2\pi\sinh\omega
    t)^{2}}\ \exp-\frac\omega 2\lbt\frac{\cosh Bt}{\sinh\omega
    t}(x-y)^{2}\right.\label{eq:lszprop}\\
&\left.+\frac{\cosh\omega t-\cosh Bt}{\sinh\omega
    t}(x^{2}+y^{2})+\imath\frac{\sinh Bt}{\sinh\omega t}x\theta^{-1} y\rbt.\nonumber
\end{align}
Note that the sliced version of (\ref{eq:lszprop}) obeys the same bound
(\ref{eq:propbound-phi4}) as the $\varphi^{4}$ propagator. Moreover the additional
oscillating phases $\exp\imath x\theta^{-1} y$ are of the form $\exp\imath\, u_{l}\theta^{-1} v_{l}$.
Such terms played no role in the power counting of the $\Phi^{4}$ theory. They
were bounded by one. This allows to conclude that Lemmas \ref{crudelemma} and \ref{improvedbound} hold
for the generalized LSZ model. Note also that in this case, the theory
contains only orientable graphs due to the use of complex fields.

\subsection{Renormalization}
\label{lszrenorm}

As for the noncommutative $\Phi^{4}$ theory, we only need to renormalize the
planar ($g=0$) two and four-point functions with only one external face.\\
Recall that the oscillating factors of the propagators are
\begin{equation}
  \label{ eq:osc-prop}
  \exp\imath\frac{\sinh Bt}{2\sinh\omega t}u_{l}\theta^{-1} v_{l}.
\end{equation}
After resolving the $v_{\ell},\,\ell\in T$ variables in terms of $X_{\ell}$,
$W_{\ell}$ and  $U_{\ell}$, they can be included in the vertices oscillations
by a redefinition of the $Q$, $S$ and $R$ matrices (see
(\ref{eq:4pt-ini})). For the four-point function, we can then perform the same
Taylor subtraction as in the $\Phi^{4}$ case.\\
The two-point function case is more subtle. Let us consider the generic
amplitude
\begin{eqnarray}
  A(G^{i}_{k})&=&\int dx dy 
  \bar{\varphi}^{\les e}(x)\varphi^{\les e}(y) \delta\big(x-y + {\mathfrak U}\big)
  \\
  &&\prod_{l \in G^{i}_{k},\, l\not\in T}du_{l}dw_{l} C_{l}(u_l, w_l)
  \nonumber\\
  &&\prod_{\ell\in T^{i}_{k}}du_{\ell}C_{\ell}(u_\ell,X_\ell,U_\ell,W_\ell) 
  \ e^{\imath XQU+\imath URU+\imath USW}\, .\nonumber
\end{eqnarray}
The symmetrization procedure
(\ref{eq:2pt-sym}) over the external fields is not possible anymore, the
theory being complex. Nevertheless we can decompose
$\bar{\varphi}(x)\varphi(y)$ in a symmetric and an anti-symmetric part:
\begin{align}
  \bar{\varphi}(x)\varphi(y)=&\,\frac
  12\lbt\bar{\varphi}(x)\varphi(y)+\bar{\varphi}(y)\varphi(x)+\bar{\varphi}(x)\varphi(y)-\bar{\varphi}(y)\varphi(x)\rbt\nonumber\\
  \defi&\,\lbt{\cal S}+{\cal A}\rbt\bar{\varphi}(x)\varphi(y).
\end{align}
The symmetric part of $A$, called $A_{s}$, will lead to the same
renormalization procedure as the $\Phi^{4}$ case. Indeed,
\begin{align}
  {\cal S}\bar{\varphi}(x)\varphi(y)=&\,\frac
  12\lbt\bar{\varphi}(x)\varphi(y)+\bar{\varphi}(y)\varphi(x)\rbt\nonumber\\
  =&\,\frac 12\lb\bar{\varphi}(x)\varphi(x)+\bar{\varphi}(y)\varphi(y)-\lbt\bar{\varphi}(x)-\bar{\varphi}(y)\rbt\lbt\varphi(x)-\varphi(y)\rbt\rb
\end{align}
which is the complex equivalent of (\ref{eq:2pt-sym}).\\
In the anti-symmetric part of $A$, called $A_{a}$, the linear terms
$\bar{\varphi}\nabla\varphi$ do not compensate:
\begin{align}
  {\cal A}\bar{\varphi}(x)\varphi(y)=&\,\frac
  12\lbt\bar{\varphi}(x)\varphi(y)-\bar{\varphi}(y)\varphi(x)\rbt\nonumber\\
  =&\,\frac
  12\Big(\bar{\varphi}(x)(y-x)\cdot\nabla\varphi(x)-(y-x)\cdot\nabla\bar{\varphi}(x)\varphi(x)\nonumber\\
    &+\frac 12\bar{\varphi}(x)((y-x)\cdot\nabla)^{2}\varphi(x)-\frac
  12((y-x)\cdot\nabla)^{2}\bar{\varphi}(x)\varphi(x)\nonumber\\
  &+\frac
  12\int_{0}^{1}ds(1-s)^{2}\bar{\varphi}(x)((y-x)\cdot\nabla)^{3}\varphi(x+s(y-x))\nonumber\\
  &-((y-x)\cdot\nabla)^{3}\bar{\varphi}(x+s(y-x))\varphi(x)\Big).\label{eq:taylor-as}
\end{align}
We decompose $A_{a}$ into five parts following the Taylor expansion (\ref{eq:taylor-as}):
\begin{align}
  A_{a}^{1+}&=\int dxdy\,\bar{\varphi}(x)(y-x)\cdot\nabla\varphi(x)\delta\big(x-y + {\mathfrak U}\big)
  \\
  &\prod_{l \in G^{i}_{k},\, l\not\in T}du_{l}dw_{l} C_{l}(u_l, w_l)
  \nonumber\\
  &\prod_{\ell\in T^{i}_{k}}du_{\ell}C_{\ell}(u_\ell,X_\ell,U_\ell,W_\ell) 
  \ e^{\imath XQU+\imath URU+\imath USW}\nonumber\\
  &=\int dx\,\bar{\varphi}(x)\,\mathfrak{U}\cdot\nabla\varphi(x)
  \prod_{l \in G^{i}_{k},\, l\not\in T}du_{l}dw_{l} C_{l}(u_l, w_l)
  \nonumber\\
  &\prod_{\ell\in T^{i}_{k}}du_{\ell}C_{\ell}(u_\ell,X'_\ell,U'_\ell,W_\ell) 
  \ e^{\imath XQ'U+\imath URU+\imath USW}\nonumber
\end{align}
where we performed the integration over $y$ thanks to the delta function. The
changes have been absorbed in a redefinition of $X_{\ell}$, $U_{\ell}$ and
$Q$. From now on $X_{\ell}$ (and $X$) contain only $x$ (if $x$ is hooked to the branch
$b(l)$) and we forget the primes for $Q$ and $U_{\ell}$. We expand the
function $f$ defined in (\ref{eq:f}) up to order 2: 
\begin{align}
  A_{a}^{1+}&=\int\bar{\varphi}(x)\,\mathfrak{U}
  \cdot\nabla\varphi(x)\prod_{l \in G^{i}_{k},\, l\not\in T}du_{l}dw_{l} C_{l}(u_l, w_l)
  \nonumber\\
  &\prod_{\ell\in T^{i}_{k}}du_{\ell}C_{\ell}(u_\ell,U_\ell,W_\ell) 
  \ e^{\imath URU+\imath USW}\nonumber\\
  &\lbt f(0)+f'(0)+\int_{0}^{1}dt\,(1-t)f^{''}(t)\rbt.
\end{align}
The zeroth order term vanishes thanks to the parity of the integrals with
respect to the $u$ and $w$ variables. The first order term contains 
\begin{align}
  \bar{\varphi}(x)\,\mathfrak{U^{\mu}}\nabla_{\mu}\varphi(x)\lbt\imath XQU+\mathfrak{R}'(0)\rbt.
\end{align}
The first term leads to
$(\mathfrak{U}^{1}\nabla_{1}+\mathfrak{U}^{2}\nabla_{2})\varphi(x^{1}U^{2}-x^{2}U^{1})$
with the same kind of expressions for the two other dimensions. Due to the
odd integrals, only the terms of the form
$(U^{1})^{2}x^{2}\nabla_{1}-(U^{2})^{2}x^{1}\nabla_{2}$ survive. We are left with integrals like
\begin{align}
\int&(u_{\ell}^{1})^{2}\prod_{l \in G^{i}_{k},\, l\not\in T}du_{l}dw_{l} C_{l}(u_l, w_l)
\prod_{\ell\in T^{i}_{k}}du_{\ell}C_{\ell}(u_\ell,U_\ell,W_\ell) 
\ e^{\imath URU+\imath USW}\; .
\end{align}
To prove that these terms give the same coefficient (in order to reconstruct a
$x\wedge \nabla$ term), note that, apart from the $(u_{\ell}^{1})^{2}$, the involved integrals are actually invariant
under an overall rotation of the $u$ and $w$ variables. Then by performing
rotations of $\pi/2$, we prove that the counterterm is of the form of the
Lagrangian. The $\mathfrak{R}'(0)$ and the remainder term in $A^{1+}_{a}$ are
irrelevant.\\
\\
Let us now study the other terms in $A_{a}$.
\begin{align}
  A_{a}^{1-}&=-\int dx\,\mathfrak{U}\cdot\nabla\bar{\varphi}(x)\,\varphi(x)
  \prod_{l \in G^{i}_{k},\, l\not\in T}du_{l}dw_{l} C_{l}(u_l, w_l)
  \nonumber\\
  &\prod_{\ell\in T^{i}_{k}}du_{\ell}C_{\ell}(u_\ell,X_\ell,U_\ell,W_\ell) 
  \ e^{\imath XQU+\imath URU+\imath USW} \; .
\end{align}
Once more we decouple the external variables form the internal ones by Taylor
expanding the function $f$. Up to irrelevant terms, this only doubles the
$x\wedge\nabla$ term in $A_{a}^{1+}$.
\begin{align}\label{thelastone}
A_{a}^{2+}&=\frac 12\int\bar{\varphi}(x)\,(\mathfrak{U}\cdot\nabla)^{2}
\varphi(x)\prod_{l \in G^{i}_{k},\, l\not\in T}du_{l}dw_{l} C_{l}(u_l, w_l)\\
&\prod_{\ell\in T^{i}_{k}}du_{\ell}C_{\ell}(u_\ell,U_\ell,W_\ell) 
\ e^{\imath URU+\imath USW}\lbt f(0)+\int_{0}^{1}dt\,f^{'}(t)\rbt.
\nonumber
\end{align}
The $f(0)$ term renormalizes the wave-function. The remainder term in (\ref{thelastone})
is irrelevant. $A_{a}^{2-}$ doubles the $A_{a}^{2+}$ contribution. Finally the
last remainder terms (the last two lines in (\ref{eq:taylor-as})) are irrelevant
too. This completes the proof of the perturbative renormalizability of the LSZ
models.\\
\\
Remark that if we had considered a real theory with a covariant derivative
which corresponds to a neutral scalar field in a magnetic background, the angular momentum
term wouldn't renormalize. Only the harmonic potential term would. 
It seems that the renormalization ``distinguishes''  the true theory 
in which a \emph{charged} field should couple to a magnetic field. It would be 
interesting to study the renormalization group flow of these kind of models 
along the lines of \cite{GrWu04-2}.

\bibliographystyle{utphys}
\bibliography{biblio-articles,biblio-books}

\end{document}